\documentstyle[11pt, aas2pp4]{article}    

\def\sqig{$\sim$}

\def\degrees{$^{\circ}$}
\def\Mesz{M\'esz\'aros}

\begin{document}

\title{
The Use of Gamma-ray Bursts as Direction and Time Markers in SETI
Strategies}


\author{Robin H. D. Corbet\altaffilmark{1}}
\affil{corbet@lheamail.gsfc.nasa.gov,\\
Laboratory for High Energy Astrophysics,\\
Code 662, NASA/Goddard Space Flight Center, Greenbelt, MD 20771}

\altaffiltext{1}{Universities Space Research Association}

\begin{abstract}
When transmitting a signal over a large distance it is more efficient
to send a brief beamed signal than a continuous omni-directional
transmission but this requires that the receiver knows where and when
to look for the transmission. For SETI, the use of various natural
phenomena has previously been suggested to achieve the desired
synchronization.  Here it is proposed that gamma-ray bursts may well
the best ``synchronizers'' of all currently known phenomena due to
their large intrinsic luminosities, high occurrence rate, isotropic sky
distribution, large distance from the Galaxy, short duration, and easy
detectability.  For targeted searches, precise positions for gamma-ray
bursts are required together with precise distance measurements to a
target star. The required burst position determinations are now
starting to be obtained, aided in large part by the discovery of
optical afterglows. Good distance measurements are currently available
from Hipparcos and even better measurements should be provided by
spacecraft now being developed. For non-targeted searches, positional
accuracies simply better than a detector's field of view may suffice
but the time delay between the detection of a gamma-ray burst and the
reception of the transmitted signal cannot be predicted in an obvious
way.

\end{abstract}

\section{Introduction}

If it is desired to transmit a signal across the Galaxy so that another,
unknown, recipient may detect it there are two basic types of
transmission patterns, an omnidirectional signal that may be detected
anywhere, or a beamed signal that can only be detected by those in the
beam. Similarly, in the time domain, a signal may either be transmitted
continuously or, with the same energy expenditure, a more powerful
signal may be transmitted for a shorter period of time.  Providing that
the recipient knows where and when a signal is coming from,
a beamed brief, and hence stronger, signal would be easier to detect.
However, for such a transmission scheme to be feasible, the problem is for
a transmitter and a recipient, one or both unknown to the other,
to find a strategy that will enable the transmitter and receiver
to transmit and observe at the right time and location.

A strategy to achieve transmitter/receiver synchronization that has been considered by
a number of authors is to utilize natural astronomical events,
see, for example, Pace \& Walker (1975), Tang (1976),
McLaughlin (1977), Makovetskii (1978), Pace (1979),
Gruber \& Pfleiderer (1982), Tang (1981),
Siebrand (1982), and Lemarchand (1994). In the
simplest scheme omnidirectional signals would be transmitted at the
occurrence of some particular event such as a nova outburst, 
maximum flux of a long period
variable, specific binary phase or supernova occurrence. A signal would then be
detected at the Earth delayed by a time corresponding to
the difference between the event/Earth distance and the
event/transmitter + transmitter/Earth distances.
The time delay is thus given by:
\begin{equation}
\Delta T = (R_s - D + (R_s^2 + D^2 - 2R_sD cos \theta )^{1/2})/c
\end{equation}
where $R_s$ is the distance to the synchronizing
astrophysical event, $D$ is the
distance to the transmitter, and $\theta$ is the angular separation
as viewed from the Earth.

A further refinement is to transmit in a direction exactly or approximately
away from the event which both reduces the time difference
and gives a preferred direction in space.
The use of one such locally
dramatic event in particular, SN 1987A, is considered by Lemarchand
(1994). The increased probability of detecting a signal if synchronizers
are used is considered by McLaughlin (1977).

To date there has been no definite detection of a signal
in the Search for Extra-Terrestrial Intelligence (SETI).
However, there have been a few detections of non-repeating
signals that
have generated some interest such as the ``wow" signal
found at Ohio State Radio Observatory (Dixon 1985,
Gray 1994)
and the
strongest events from the META survey which appear to preferentially
lie in the Galactic plane (Horowitz \& Sagan 1993). While
these may simply be noise or arise from natural astrophysical phenomena
they could conceivably be genuine extra-terrestrial artificial signals that
are transient either because transmission is intermittent or caused
by interstellar scintillation (Cordes, Lazio, \& Sagan 1997).

In this paper the use of one particular type of natural
synchronizing signal is
considered - the phenomenon of gamma-ray bursts (GRBs). These appear to
posses a number of important advantages over other possible astrophysical
events and
their use in SETI is advocated for, in particular,
targeted observations of relatively nearby stars. A brief summary of the phenomenology of
gamma-ray bursts and their observations is given in Section 2, followed
in Section 3 
by a
brief comparison of the use of GRBs to some other
possible synchronizers,
and in Section 4 
two possible strategies for utilizing
gamma-ray bursts in extraterrestrial communication
are discussed.

\section{Gamma-ray Bursts}

GRBs are found to occur istrotropically over the sky (e.g. Briggs
1993)
and the BATSE detector on board the Compton Gamma-Ray Observatory
(CGRO) detects approximately one burst per day. GRBs are typically
rather
rapid events with log(T90), where T90 is the time within which 90\% of the flux
from a burst is contained, showing a bimodal
distribution with peaks at about 0.3 and 20 seconds (Fishman
et al. 1994).
Afterglows in the X-ray, optical, and radio bands have also recently
been detected (see e.g. the review by \Mesz\ 1998)
and the optical and X-ray afterglows have been seen to decay away
with power-law indices of between roughly 1 to 2.

The GRB intensity distribution shows that brighter bursts follow
a peak flux distribution with a power-law index
of ${-3/2}$  as expected for a uniform spatial distribution
(e.g. Fenimore et al. 1993). However, fainter bursts show a flattening
curve for which a simple interpretation is that we are observing
the ``edge'' of the GRB distribution. This
interpretation is complicated though by the unknown range of luminosities
that GRBs may display.

Although GRBs were once considered to arise from phenomena arising on
Galactic neutron stars, the detection within the last year of so of a
number of afterglows associated with host galaxies has enabled their
distances to be firmly established as ``cosmological'', i.e. at very
large distances. For the first three cases where a redshift was
measured, values of Z = 0.835, 3.418 and 0.966 were obtained for GRBs
970508, 971214 and 980703 respectively (Metzger et al. 1997, Kulkarni
et al. 1998, Djorgovski et al. 1998). At these large distances the
implied luminosities of GRBs are extremely large. For example, the
implied isotropic gamma-ray luminosity of GRB 971214 is
\sqig3$\times$10$^{53}$ ergs s$^{-1}$ (Kulkarni et al. 1998).  Even if
the probable significant beaming is taken account of, the energy
involved is at least comparable to, and likely exceeds, that associated
with supernovae.

The astrophysics of gamma-ray bursts is currently very poorly
understood and current explanations to account for GRBs include merging
neutron stars (e.g. \Mesz\ 1998) and hypernovae (Paczy\'nski 1998).
However, the mechanism that causes GRBs is {\em not} important for
their use in synchronizing communication across large distances.

Although the BATSE detector is more sensitive
than earlier generations of instruments it has only limited precision
in locating bursts and and even bright bursts are not positioned to
much better than a few degrees (Pendleton et al. 1999). Progress in
precisely locating GRBs has recently come from the instruments onboard
the SAX satellite. In addition to a GRB detector, SAX
carries X-ray instrumentation consisting of a Wide-Field
Camera (WFC) and a suite of Narrow Field instruments. For those bursts
that occur within the field of view of the WFC, a position accurate to 3
to 8 arc minutes can be obtained, follow-up observations
of X-ray afterglows with the
narrow field instruments several hours later can then yield positions
to about one arcminute (e.g. In't Zand et al. 1998).  The All Sky
Monitor on-board the Rossi X-ray Timing Explorer has also provided
some locations for GRBs to a few arc minutes (Smith et al. 1998).
It is this provision of arc minute accuracy positions that has
lead to the discovery of the optical afterglows which can then
make it possible to obtain positions accurate to better than an
arc second.

In the
near future, the HETE-II mission (Ricker 1997) should provide positions
accurate to 10 arc seconds to arc minutes for about 30 bursts per
year.  A further possible future GRB mission is Swift 
(Gehrels et al. 1999). Swift is designed to produce positions accurate to better
than an arc second (if optical emission is also detected) within better
than 90 seconds and would view an area of 2 steradians.  The
sensitivity of the Swift GRB detector is such that it is expected that
it will detect at least 300 bursts per year. If funded it is
intended that Swift would be launched in 2003.

Systems of GRB detectors in spacecraft spread across the solar system
in interplanetary networks (see e.g. Hurley et al. 1994) can also
provide locations by comparing the time of arrival of a burst at the
various spacecraft.  Generally this type of network has not so far
given very rapid determinations of source location, this can be caused
by, for example, infrequent contacts with the spacecraft from ground
stations or the difficulties of merging data obtained with a variety
instruments in possibly differing formats.

\section{Comparison with Some Other Astrophysical Synchronizers}

Probably the most important property of a synchronizer is
that it should, in some sense, be obvious and so likely to be
used by both the transmitter and receiver.
Arguably the key property
in making a phenomenon obvious is its luminosity. In addition,
it is desirable for 
the phenomenon to occur at a sufficiently rapid rate
to facilitate the use of a large number of these events. Further,
it should be possible to determine their time of occurrence
precisely to make it possible to use a brief artificially 
transmitted signal.

After GRBs, supernovae are the next most energetic phenomena known in the
universe. However, it is considerably more difficult to detect
all supernovae down to a specified flux limit than is the
case for GRBs.
The positions of supernovae are typically derived from optical
observations and are thus known to high precision.
However, the optical light curves of supernovae are relatively slowly
varying and so are significantly less sharp time markers.
Supernovae could
perhaps be of better use if larger numbers could have their exact times
of onset measured via, for example,  a neutrino pulse (cf. SN 1987A
Bionata et al. 1987, Hirata et al. 1987).
Note that it has
been suggested that a small subset of GRBs may be caused by the unusual
Type Ic supernova class based on a possible association
between SN1998bw, which had a redshift
of only 0.008, and GRB980425 (e.g. Kippen et al. 1998).

Novae and other variable stars have also been proposed as possible
synchronizers.  Novae are significantly less energetic than supernovae
and,
in the case of Galactic novae, it is also typically necessary to
accurately know the distance to the nova as well as the
transmitter to be able to calculate time delays.  Variable stars form
a broad diverse ``class'' of objects.  Their much lower luminosities
and varied nature makes them less obvious as the synchronizers that
would be universally used.

Until recently, the relatively poor precision with which gamma-ray
burst locations were determined was arguably a problem with their use
as synchronizers as this yields a large uncertainty in the time delay
(eqn. 1).  However, this limitation is disappearing with new
generations of instruments augmented by the discovery of, in
particular, optical afterglows. As the most energetic of all currently
known natural phenomena this alone draws attention to GRBs. They occur
at rapid rates, one a day even with BATSE level technology, are easily
detectable, have short durations making them ``sharp'' time markers,
and also occur at sufficiently large distances that it is not necessary
to actually know these distances to calculate time delays.
Hence, no other class of proposed phenomenon appears to posses any
obvious properties which makes it a better candidate than GRBs for use
as a synchronizer.

\section{Use of GRBs for SETI}

It is now considered how GRBs may be used as synchronizers in SETI and
two basic transmission strategies are investigated:

(i) Targeted Signaling. In this case the transmitter sends a signal
promptly after the detection of a GRB in one or more directions
``close'' (i.e. within an angle $\theta$) to the opposite direction of
the GRB at target(s) previously decided to be of interest. The beam
width is likely to be made as small as possible.

(ii) All-sky signaling. The transmitter sends a signal promptly
after the detection of a GRB in a direction exactly opposite
from the GRB with a beam half-width $\theta$.

For a GRB located at a very large distance from both the transmitter
and receiver, as appropriate for a GRB,
the time delay between the detection of a GRB and the prompt transmitted
signal is given by the simplified expression:\\
$\Delta T = (1 - cos\theta) D/c$

For small angles $\Delta T = D \theta^2/c$, for $\theta$
in radians. Note that $D$ and $\theta$ formally refer to the
distance and angular separation when the burst arrives
at the transmitter rather than the receiver.

\subsection{Targeted Signaling}

It is not possible to know in advance the angle within which the
transmitter will decide to broadcast to particular locations downstream
from the GRB.
The smaller the angle utilized the longer it will take
to transmit to all targets. However, for large angles
the time delays will be larger and harder to accurately calculate for
the receiver and more targets may have to be transmitted
to for any particular GRB.
A non-exhaustive list of
factors which could be used by the transmitter to
decide which places to target, in likely order of the distance
at which they could be detected by the transmitter, is:
\begin{itemize}
\item{Detection of a solar type star}
\item{Detection of a planetary system with Jupiter mass planet(s)}
\item{Detection of terrestrial mass planet(s)}
\item{Detection of life through the presence of, for example,
an Oxygen atmosphere}
\item{Detection of intelligent life through, for example, 
radio transmissions}
\end{itemize}

The down-stream angle(s) used by a transmitter will presumably depend
on the number of targets that are thought to be potential hosts for
receivers for the signal. For a large number of possible targets small
angles would perhaps be likely to be used.  Conversely, at the small
number extreme, the Earth would be the only target transmitted to even
if the Earth was exactly upstream from the GRB event.

In order for the receiver to be able to calculate the time delay from a
potential transmitter it is necessary
to know the distance to that transmitter
and the angular distance from the GRB.  The current best set of stellar
distances comes from HIPPARCOS parallax measurements (Perryman et al.
1997) which gave values accurate to \sqig 1 milli-arcsecond. In the the
near future NASA's Space Interferometry Mission (SIM, Unwin et al.
1998) and ESA's GAIA (Gilmore et al.  1998) may yield 
large numbers of parallaxes with
precisions better than \sqig 10 micro-arcseconds.

For an illustration of the resulting
time delays, and the errors involved in
calculating these, the three classes of objects in the Project Phoenix
targeted survey are considered (Henry et al. 1995) and presented in
Table 1. Time delays are given for the three maximum
distances corresponding to the
``Nearest 100'' (D $<$ 7.2 pc), ``Best \& Brightest'' (D $<$ 20 pc),
and ``G Dwarf'' (D $<$ 50pc) targets
as well as for distances of 100 pc and 1000 pc.  Errors
in calculating time delays due to uncertainties
in distance measurements from HIPPARCOS
and a future GAIA/SIM class mission are listed. Time delay errors due
to GRB position determinations accurate to 1\arcsec\ and 10\arcsec\ are
also given as appropriate to future GRB missions.  For the closest
class it appears feasible to use even 10\arcsec\ accuracy GRB locations
and HIPPARCOS distances for moderately large offset angles. For the
largest distances considered in this table of 1000 pc, at angles of
only 1\degrees\ even parallax measurements accurate to 10
micro-arcsecond yield uncertainties in the arrival of a signal of almost
2 days. For all three of these Project Phoenix target classes
GAIA/SIM class parallax measurements combined with good GRB
measurements give errors on time delays that are modest ($<$ 1 day)
for offset angles up to 5\degrees.

\subsection{All-sky Signaling}

If the intention is to eventually broadcast to the entire
sky with no preference for particular
locations there are two competing constraints:\\
(i) The use of a narrow beam width will produce both a larger gain and
also reduce the maximum possible time delay at the receiver between the
detection of a GRB and the transmitted signal. \\
(ii) The narrower the beam the longer it will take to cover the entire sky.

One consideration may lead to a ``natural'' beam width for
transmissions.  If the transmitter wishes to broadcast out to a certain
distance $D$, then a beam width can be chosen such that the maximum
time delay at $D$ would be equal to the average interval between GRBs
for the intensity level that they are using to select GRBs.
i.e.
$
\Delta T = 1 / R
$
where $R$ is the GRB/transmission rate.
This yields
$
\theta\ = (c/DR)^{1/2}
$

Note that basing a beam width on this consideration means that using
additional lower intensity GRBs would not alter the time taken to
illuminate the entire sky as the time delay and area of sky illuminated
both depend on $\theta^{2}$. However, using additional bursts does result
in a narrower beam and hence larger gains and smaller maximum
time delays.
The assumed transmission distance, $D$
might vary depending on the transmission direction, for example
transmissions perpendicular or parallel to the Galactic plane. In
addition, the beam width could potentially also be altered depending on
the luminosity of a gamma-ray burst so that the maximum time delay at a
certain distance would be the mean recurrence time for bursts of that
luminosity or greater.

The time taken to illuminate an area equal to the entire sky,
although with overlap,
is given by
$
T = 4 \pi / (R  B)
$
where B is the area of the beam in steradians.
Hence the ``natural'' beam width gives $T = 4D/c$.
Thus, for any significant distance, a considerable time is
required to illuminate the entire sky.
Alternatively, if this ``natural'' beam width is not used then
the number of GRB locations that needs to be monitored is
$\sim R  \Delta T$ = $ R  D  \theta^2/c $.

\section{Conclusion}

Gamma-ray bursts posses a number of properties that make them very good
candidates for synchronizers that could aid in the search for brief
beamed extraterrestrial signals. GRB positions are now starting to be
obtained with sufficient precision to make them useful for targeted
searches. This is augmented by precise measurements of stellar
parallaxes from satellite borne instruments, and it is anticipated that
even better parallax measurements will be obtained in the relatively
near future.  GRB locations and times are rapidly available from the
GCN (GRB Coordinates Network, Barthelmy et al. 1998) and this
is likely to continue with future missions which also aids in their
use.

For all sky surveys, the case for using GRBs as target positions to
observe may be a lot weaker. However, there appears to be little to
lose by pointing a detector as soon as possible at GRB locations and
monitoring that location for some time.  Additionally, even the
positions currently provided by BATSE may be used for this purpose by a
detector that has a field of view of a few degrees.

\begin{table}
\caption{Time Delays between GRB and Transmitted Signal}
\begin{center}
\begin{tabular}{llccccc}
Distance (pc) & Angle(\degrees) & Delay (d) & 1\arcsec & 10\arcsec & Hipparcos & GAIA/SIM \\
\tableline
7.2 & 1 & 1.30631 &   7.25609E-04 &   7.26718E-03 &   9.47365E-03 &  9.40612E-05\\
7.2 & 2 & 5.22485 &  1.45110E-03 &  1.45221E-02 &  3.78917E-02 &  3.76216E-04\\
7.2 & 3 & 11.7544 &  2.17615E-03 &  2.17725E-02 &  8.52456E-02 &  8.46379E-04\\
7.2 & 5 & 32.6379 &  3.62403E-03 &  3.62514E-02 & 0.236697 &  2.35010E-03\\
\tableline
20 & 1 & 3.62864 &  2.01558E-03 &  2.01866E-02 &  7.40539E-02 &  7.25874E-04\\
20 & 2 & 14.5135 &  4.03083E-03 &  4.03391E-02 & 0.296193 &  2.90327E-03\\
20 & 3 & 32.6512 &  6.04485E-03 &  6.04793E-02 & 0.666350 &  6.53154E-03\\
20 & 5 & 90.6608 &  1.00668E-02 &  1.00698E-01 &  1.85022 &  1.81358E-02\\
\tableline
50 & 1 &9.07161 &  5.03895E-03 &  5.04665E-02 & 0.477453 &  4.53807E-03\\
50 & 2 & 36.2837 &  1.00771E-02 &  1.00848E-01 &  1.90967 &  1.81509E-02\\
50 & 3 & 81.6279 &  1.51121E-02 & 0.151198 &  4.29621 &  4.08344E-02\\
50 & 5 & 226.652 &  2.51669E-02 & 0.251746 &  11.9291 & 0.113383\\
\tableline
100 & 1 & 18.1432 &  1.00779E-02 &  1.00933E-01 &  2.01591 &  1.81614E-02\\
100 & 2 & 72.5673 &  2.01541E-02 & 0.201695 &  8.06304 &  7.26400E-02\\
100 & 3 & 163.256 &  3.02242E-02 & 0.302396 &  18.1395 & 0.163419\\
100 & 5 & 453.304 &  5.03338E-02 & 0.503491 &  50.3671 & 0.453758\\
\tableline
1000 & 1 & 181.432 &  1.00779E-01 &  1.00933 &  --- &  1.83265\\
1000 & 2 & 725.673 & 0.201541 &  2.01695 & --- &  7.33003\\
1000 & 3 & 1632.56 & 0.302242 &  3.02396 & --- &  16.4905\\
1000 & 5 & 4533.04 & 0.503338 &  5.03491 & --- &  45.7883\\
\tableline

\tableline
\end{tabular}
\end{center}
Notes: Time delay and errors are in days. Angle is the
offset between the gamma-ray burst and the transmitter.
The 1\arcsec\ and 10\arcsec\ columns list
the uncertainties due to inaccuracies of 1 and 10\arcsec\ in
the position of the GRB respectively. The HIPPARCOS and GAIA/SIM columns
give the errors due to uncertainties in the parallax of
a transmitter of 1 milli-arcsecond and 10 micro-arcseconds respectively.
\end{table}

\end{document}